\documentclass[preprint]{aastex}
\usepackage{epsf,epsfig}

\shorttitle{The Red Clump as a Distance Indicator}
\shortauthors{Grocholski and Sarajedini}

\begin{document}
    
\def\lea{\mathrel{<\kern-1.0em\lower0.9ex\hbox{$\sim$}}}
\def\gea{\mathrel{>\kern-1.0em\lower0.9ex\hbox{$\sim$}}}

\title{WIYN Open Cluster Study. X. The K-Band Magnitude \\ of the
Red Clump as a Distance Indicator}

\author{Aaron J. Grocholski and Ata Sarajedini}
\affil{Department of Astronomy, University of Florida, 
P. O. Box 112055, Gainesville, FL 32611  
\\ aaron@astro.ufl.edu, ata@astro.ufl.edu}

\begin{abstract}
In an effort to improve the utility of the helium burning red clump 
luminosity as a distance indicator, we explore the sensitivity of the 
K-band red clump absolute magnitude [$M_K(RC)$] to metallicity and age. 
We rely upon JK photometry for 14 open clusters and two globulars 
from the 2nd Incremental Data Release of the 2MASS Point 
Source Catalog. The distances, metallicities, and ages of the open 
clusters are all on an internally consistent system, while the K(RC) values 
are measured from the 2MASS data. For clusters younger than 
$\sim$2 Gyr, $M_K(RC)$ is insensitive to metallicity but shows a 
dependence on age. In contrast, for clusters older than $\sim$2 Gyr,
$M_K(RC)$ is influenced primarily by the metallicity of the population 
and shows little or no dependence on the age. Theoretical red clump 
models based on the formalism of Girardi et al. reinforce this finding.
Over comparable metallicity and age ranges, our average $M_K(RC)$ 
value is in accord with that based on solar-neighborhood red clump stars
with HIPPARCOS parallaxes. Lastly, we compute the distance to 
the open cluster NGC 2158 using our red clump calibration. Adopting an age 
of $1.6 \pm 0.2$ Gyr and $[Fe/H] = -0.24 \pm 0.06$, our calibration yields a 
distance of $(m-M)_{V} = 14.38 \pm 0.09$.
\end{abstract}

\keywords{open clusters and associations: general -- stars: 
color-magnitude diagrams -- stars: distances -- stars: horizontal 
branch}

\section{Introduction}

During the past few years, the helium burning red clump (RC) has gained
considerable attention for its potential as a standard candle. The
primary advantage of the RC is the ease with which it can be 
recognized in the color-magnitude diagram. However, there is
currently a great deal of controversy in the literature regarding
the appropriate treatment of possible metallicity and age effects on 
the I-band absolute magnitude of the RC ($M_I(RC)$).
There are two schools of thought on this issue; the first assumes a
constant value for $M_I(RC)$ which is then used to facilitate 
a single-step distance determination via knowledge of the apparent RC 
magnitude and the extinction (e.g. Paczy\'nski \& Stanek 1998;
Stanek \& Garnavich 1998). The second approach is founded on the claim
that both age and metal abundance have a significant influence on 
the luminosity of RC stars (e.g. Cole 1998; Sarajedini 1999) and 
must be accounted for
in determining $M_I(RC)$ and therefore the distance.

Both Paczy\'nski \& Stanek (1998) and Stanek \& Garnavich (1998) use
HIPPARCOS RC stars with parallax errors of less than 10\% to calculate the
I-band absolute magnitude of the solar neighborhood red clump.  
In their analysis,
Paczy\'nski \& Stanek (1998) find that $M_{I}(RC)$ shows no variation 
with color
over the range $0.8 < (V-I)_{0} < 1.4$ and, from a Gaussian fit to 
the RC luminosity function, find $M_{I}(RC) = -0.28 \pm 0.09$.  Following the
same methodology and building upon the 
earlier work, Stanek \& Garnavich (1998)
find a similar result with $M_{I}(RC) = -0.23 \pm 0.03$.
With this calibration, a single step calculation is then used to determine
the distance to the Galactic center (Paczy\'nski \& Stanek 1998) and M31
(Stanek \& Garnavich 1998).  Both of these investigations found 
little or no variation in $M_I$ of the RC stars with color;
this was taken to imply that $M_{I}(RC)$ does not vary significantly with 
metallicity.  

In contrast, theoretical models from Girardi \& Salaris (2001) and the 
earlier models of Seidel, Demarque, \& Weinberg (1987; see also Cole 
1998) show that $M_I(RC)$ is dependent on both age and metallicity, becoming
fainter as both increase.  These models are in good agreement with the
observations presented by Sarajedini (1999, hereafter S99).  Using published
photometry for 8 open clusters, S99's most important result is 
that while $M_{I}(RC)$ is less sensitive to metal abundance than 
$M_{V}(RC)$, both still retain a considerable dependence on the age 
and metallicity of the stellar population. As a result, the single-step 
method of 
applying the solar-neighborhood $M_{I}(RC)$ to populations with a 
different age-metallicity mix could be problematic.


Alves (2000) also uses the HIPPARCOS RC for his calibration; 
however, he relies upon the K-band luminosity ($M_K$) of the RC stars 
in the hope that, since the K-band is less sensitive to extinction 
(and possibly metallicity as well) than the I-band, it might make a 
better choice as a standard candle.  Alves (2000) restricts his RC stars 
to those that have metallicities from high resolution spectroscopic data.  
For this group of 238 RC stars, he finds a peak value of 
$M_K(RC) = -1.61 \pm 0.03$ 
with no correlation between $[Fe/H]$ and $M_K$. However, he is not 
able to explore the effect of age on $M_K(RC)$ due to the lack of 
such information for the individual stars in his sample.

These previous works have prompted us to combine the approaches of S99 
and Alves (2000) and to investigate the influence of age and metal 
abundance on $M_K(RC)$ for a 
number of open clusters with well-known distances and metallicities.  
In section 2 we discuss the observational data.  
Section 3 compares our data with the results of theoretical models
and presents a discussion of the results; our conclusions are 
summarized in Section 4.

\section{The Data}

\subsection{Open Clusters}

In the present study, the most important criterion that the 
observational data must fulfill is that of internal consistency.
For example, we must ensure that the distance moduli, reddenings, ages, 
and metallicities of all of the clusters in our sample have been 
determined using the same techniques. In addition, it is imperative 
that the IR photometry we rely upon be measured and calibrated 
in a consistent manner. For the former, we use the database of open 
cluster properties as measured by Twarog et al. (1997), supplemented by 
cluster ages from the WEBDA\footnote{http://obswww.unige.ch/webda/webda.html} 
database, and, for the latter, we utilize the 2nd 
Incremental Data Release of the 2MASS Point Source 
Catalog\footnote{http://irsa.ipac.caltech.edu/}. We now discuss each of 
these in more detail.

Twarog et al. (1997) have compiled 
a list of 76 open clusters for which they provide reddenings, 
distance moduli and metallicities. For the purposes of the present 
paper, we limit ourselves to distance moduli derived via the technique 
of main sequence fitting so as to remain independent of methods that 
rely on the luminosity of the RC. Their metallicities have all been
measured on the same system and the reddenings have been determined
using an internally consistent method.  The vast majority of these values 
are consistent with those found in the literature, 
except for the reddening of NGC 6819 for which the Twarog et al.
(1997) value is much higher than other published values.  As a result,
we have decided to adopt the S99 reddening for NGC 6819 
instead of the apparently discrepant value tabulated by Twarog et al.
(1997). Additionally, because the determination of the reddening and 
distance modulus are coupled, we also adopt the S99 distance modulus for 
NGC 6819.  

The ages of the open clusters have been obtained from WEBDA, which 
is a compilation of
open cluster data from various sources.  To check the
reliability of the WEBDA ages, we compare them with the cluster
ages determined by S99 in Fig. 1.  S99 presents 
isochrone-fitting ages for 8 open clusters that have been determined 
in the same manner using the Bertelli et al. (1994) theoretical 
isochrones.
The left panel of Fig. 1 plots the ages from S99 versus the
ages in WEBDA, where both axes are in log space and the
dashed line represents a zero age difference between the systems.  From
this plot, it is evident that there is a systematic offset between the two
systems with the WEBDA ages being younger than those of S99.  The
average difference, $\Delta$$log (Age) = 0.191$, is used to shift the 
ages given in WEBDA onto the S99 system.
The right panel of Fig. 1 shows the ages from S99 plotted against the
shifted WEBDA ages; it is clear from Fig. 1 that the shifted
ages are in better agreement with those of S99; as a result,
we will apply this shift to nine of the clusters in our study
and use the S99 ages for the five clusters that are common to both studies.  

For the open clusters in the Twarog et al. (1997) study that possess main 
sequence fitting distances, we extracted $JHK$ photometry from the 2nd 
Incremental Data Release of the 2MASS Point Source Catalog.
As noted above, these data have been obtained using 
similar instruments and reduced with the same pipeline techniques.
For each cluster,
we have utilized the same criteria for the 2MASS data retrieval.  The 
field size is originally set to 30 arcmin in radius 
and then reduced to fields as small as 5 arcmin in radius 
in an attempt to isolate the cluster stars.  
The sources are limited to a brightness of 6th magnitude 
or fainter due to saturation effects at the bright end (Carpenter 2000).  
Lastly, we have extracted only the highest quality photometry from 
the 2MASS catalog, which provides a read flag (rd\_flg) indicating
how the photometry of each star was measured.  We have chosen to 
exclude any source 
that has a read flag of zero in any band since this implies that the 
source was not detected in that band and the magnitude given is an upper
limit. We note that the vast majority of the stellar magnitudes used 
in this study are based on point-spread-function fitting 
(i.e. rd\_flg = 2); however, in order to include the brighter red clumps
of more nearby clusters, we have had to use the aperture photometry in a 
minority of cases.

The 2MASS program uses a K-short ($K_S$) filter for their 
observations. We have chosen to convert these magnitudes to the K-band
adopting the Bessell \& Brett (1988, hereafter BB) system, which is 
also used in the theoretical 
models of Girardi et al. (2000) and Girardi \& Salaris (2001).  The 
transformation equations 
are derived by Carpenter (2001) and are adopted as follows:
\begin{equation}
(J-K_{BB}) = [(J-K_S) - (-0.011 \pm 0.005)] / (0.972 \pm 0.006)
\end{equation}
and 
\begin{equation}
K_{BB} = [K_S - (-0.044 \pm 0.003)] - (0.000 \pm 0.005) * (J-K_{BB}).
\end{equation}
We note, however, that 
transformation to the Koornneef (1983) K-band (used by Alves 2000, Sec. 
3.2) would have a negligible effect on our results.  To correct for the 
interstellar reddening, we adopt the extinction law determined by 
Cardelli et al. (1989), which, using their value of $R_{V}=3.1$, gives 
$A_{K} = 0.11 A_{V}$ and $A_{J} = 0.28 A_{V}$.  From this, it is a simple 
matter to calculate the absolute K-band magnitude and dereddened $J-K$ 
color of the open cluster stars.

We have determined the RC luminosity for our clusters by taking 
the median value of $M_K$ for all stars within a standard sized box 
placed around the RC.  We use the median value of
$M_K$ along with a constant box size in an attempt to eliminate any
selection effects that may occur in choosing the location of the RC
and to limit the effect of outliers on $M_{K}(RC)$.  
Figure 2 shows the color-magnitude diagrams (CMDs) for all
14 clusters focused on the RC and main sequence turnoff.  The
box used to select the RC stars for each cluster is also shown. 
We note that, where available, we have used published optical CMDs for 
our clusters to help isolate the approximate RC location.  The 
uncertainty in $M_K(RC)$ is calculated by combining 
the standard error about the mean K-magnitude for all stars 
inside the red clump boxes along with the errors in $E(B-V)$ and 
$(m-M)_{V}$, all added in quadrature. Except where otherwise noted,
we adopt 20\% of the value as the error in $E(B-V)$ and 10\% of 
the value as the error in $(m-M)_V$.  


\subsection{Globular Clusters}

It is difficult to ensure that the distances, ages, and metallicities of 
globular clusters with RCs are on the same system as those of the 
open clusters. Fortunately, the ages and metallicities of the two globulars 
in our sample - 47 Tuc and NGC 362 - are sufficiently different 
from the bulk of the open clusters that small systematic discrepancies 
in these quantities should not be a significant hindrance to the 
interpretation of the results. In any case,  we have decided to adopt 
literature values 
for the basic cluster parameters and use the globular cluster RCs as a 
consistency check.  

In the case of 47 Tuc, we adopt the metallicity 
quoted by Carretta \& Gratton (1997) of $[Fe/H] = -0.70 \pm 0.07$, which 
happens to be very close to the Zinn \& West (1984) value. For the distance 
modulus and reddening, we average the published values listed in Table 
2 of Zoccali et al. (2001) to obtain $(m-M)_{V} = 13.45 \pm 0.21$ and
$E(B-V) = 0.044 \pm 0.008$,
where the errors represent half of the range of tabulated values.
Lastly, for the age of 47 Tuc, we adopt the oldest age for which the 
models predict the presence of a red clump at its metallicity - 12 Gyr.

For NGC 362, we adopt a similar approach. The metal abundance of
$[Fe/H] = -1.15 \pm 0.06$ is taken from Carretta \& Gratton (1997), which 
is approximately 0.1 dex more metal-rich than the Zinn \& West (1984) 
value. Our search of the literature has revealed distance moduli that 
range from $(m-M)_{V}$ = 14.49 (Zinn 1985) to 14.95 (Burki \& Meylan 
1986; see also Bolte 1987) and reddenings in the range 
$E(B-V)$ = 0.032 (Vandenberg 2000) to 0.08 (Alcaino 1976) leading to
adopted values of 
$14.70\pm0.23$ and $0.048\pm0.024$ for the apparent distance modulus and 
reddening of NGC 362, respectively. Once again, we adopt an age of 12 Gyr.

The red clumps of these globulars have 
been isolated in the 2MASS point source catalog in the same way as for 
the open clusters. The ($M_{K}, (J-K)_{0}$) CMDs for 47 Tuc and NGC
362 are shown in Fig. 3 along with the box used to define their RCs. 
All of the relevant observational parameters for the open and 
globular clusters are listed in Table 1.

\section{Results and Discussion}

\subsection{Cluster Data}

As described in Sec. 1, we are interested in
exploring the dependence of $M_K(RC)$ on $[Fe/H]$ and age.  Plotted in
Fig. 4 are the $M_K(RC)$ values for the 
14 open clusters (open circles) and the two globulars (filled circles) 
in our sample versus the logarithm of the age
(top right panel) and the metallicity (top left panel). 
The red clumps start out very
bright at young ages and decrease in brightness by almost one magnitude
as the cluster ages approach $10^9$ years after which they brighten by 
$\sim$ 0.5 magnitude.  
Then, the red clumps become slightly fainter as the cluster ages 
increase up to $10^{10}$ years.  The top two panels of Figure 4 also 
include $M_I(RC)$ (open squares) from S99.  Keeping in mind that the 
numbers of clusters is small, over the age and metallicity range common 
to both studies, the K-band absolute magnitude of the RC exhibits less 
sensitivity to age and metal abundance than does $M_{I}(RC)$.


\subsection{Field Star Data}	

Alves (2000) reports K-band absolute magnitudes of
solar-neighborhood stars in the HIPPARCOS catalog along with 
parallaxes and proper motions. Following the analysis of Alves 
(2000), we have limited ourselves to stars with
$2.2 < (V-K)_{0} < 2.5$ and $-2.5 < M_{K} < -0.8$ in Table 1 of 
Alves' paper in order to isolate a sample of nearby red clump stars.
We have plotted $M_K$ vs $[Fe/H]$ for these stars in the 
bottom panel of Figure 4 (open circles). For comparison, the open and 
globular cluster data from the present work (filled circles) are also 
shown.  Keeping in mind that there are far fewer open clusters than 
field stars in Fig. 4, we find good consistency in the locations of 
the two samples. Alves (2000) finds $\langle$$M_{K}(RC)$$\rangle$$
= -1.61 \pm 0.03$, while the open clusters in our sample give
$\langle$$M_{K}(RC)$$\rangle$$= -1.62 \pm 0.06$.
This is remarkable given the fact that the open 
cluster distances are based on the main sequence fitting results of 
Twarog et al. (1997) and the field stars are on the HIPPARCOS 
distance scale. Both show mean K-band magnitudes of $\sim$--1.6 and 
very little if any dependence on metal abundance over the same range.

\subsection{Comparison With Theoretical Models}

Leo Girardi has kindly provided us with theoretical models that 
represent the median magnitude of the red clump as a function of age 
and metal abundance (Girardi \& Salaris 2001; Crowl et al. 2001).  
Figures 5 and 6 show these theoretical models in the K-band 
compared with our open and globular cluster data.  The nine panels of 
Fig. 5 display the K-band luminosity of the red clump as a function
of metallicity for a range of ages.  The five panels in Fig. 6 
illustrate 
the variation of the K-band luminosity with age for a range of metal 
abundances. 
In both of these figures, clusters that are similar in metallicity or age 
to the model plotted in each panel are represented by filled circles while 
the remaining clusters are denoted by open circles.  
Figures 5 and 6 suggest that at ages younger than $\sim$2 Gyr, the red 
clump luminosity is greatly dependent on the age of the cluster and shows 
little effect from the metallicity whereas clusters older than $\sim$2 Gyr 
show the exact opposite, having little age dependence while still showing 
the effects of metallicity.  

To facilitate a more detailed comparison between the models and 
the observations, we utilize an interpolation routine based on 
low order polynomials to compare the theoretical $M_{K}$ values with the 
observed ones.  As a test of the interpolation, we have 
applied it to the observational data alone comparing the interpolated 
$M_K(RC)$ values to the actual values at the age and abundance of each 
cluster.
We find that the root-mean-square of the residuals is negligible in 
$M_{K}(RC)$ with no systematic trends as a function of age or 
abundance.  We 
have also tested the interpolation routine on the theoretical models 
with similar encouraging results.  
The accuracy of the interpolation allows us to compare the $M_K(RC)$ values 
predicted by the models for a given age and metallicity to the observed 
$M_K(RC)$ for each cluster.  From this comparison, we find that the 
root-mean-square deviation of the theoretical models from the observations
is 0.16 mag, with no systematic variation in the residuals as a function of 
age or metallicity. This deviation is slightly larger than the mean 
error in the $M_K(RC)$ values of the 16 clusters, which we find to be
0.13 mag. Given that the mean deviation of the models from the 
observations is roughly consistent with the errors inherent in the 
latter, it is reasonable to 
conclude that the models are generally consistent with the 
observational data.

Figure 7 shows the Girardi models plotted along with
the Alves (2000) field red clump star data.  The models reinforce the 
conclusion drawn by Alves (2000) that $M_{K}$ is insensitive to 
metallicity for nearby stars in this abundance range.  Furthermore, 
given that a typical $M_K$ error in Alves (2000) data is 0.11 mag, 
this figure suggests that the vertical spread in the $M_{K}$ values is 
mainly the result of age effects among the field stars.  Both the 
$10^{8.8}$ and $10^{9.2 - 9.4}$ year isochrones agree with the majority 
of the data. However, given the expectation that stars in the Solar 
neighborhood are 
likely to be near Solar-age, most of the HIPPARCOS stars in Fig. 7 
probably have Log ages between 9.2 and 9.6 (1.6 to 4.0 Gyr).

It is interesting to note that the ages of the solar 
neighborhood stars as predicted by the models show a lack of stars around 
$10^9$ years (Fig. 7).  In contrast, using a model of the solar 
neighborhood RC that assumes a constant star formation rate, Girardi \& 
Salaris (2001) expect an age distribution for the RC stars that peaks at 
1 Gyr with approximately 60\% of the stars having this age.  The 
discrepancy in this result with the apparent ages of the HIPPARCOS RC 
stars likely indicates a non-constant star formation rate in the solar 
neighborhood; this is not surprising if the formation of stars is 
triggered by density waves traveling through the solar neighborhood, 
which is an intrinsically episodic process.

\section{Application As a Distance Indicator}

An important aspect of this study is the application of the K-band red
clump absolute magnitude as a distance indicator. To optimize this 
application in the present work, we seek a range of age and abundance 
over which variations 
in $M_{K}(RC)$ are minimized. Inspecting Fig. 4, we see that if the 
age of the stellar population is in the range 2$\lea$Age$\lea$6 Gyr and
the metal abundance is between --0.5$\lea$[Fe/H]$\lea$0.0, then the 
intrinsic variation in $M_{K}(RC)$ is minimized suggesting that 
uncertainties in our knowledge of these properties are inconsequential 
in the determination of the distance.
Based on these considerations, we have selected the open cluster NGC 2158. 
This cluster possesses 2MASS photometry, and it is included in 
the study of Twarog et al. (1997), so we have a metallicity 
value ($[Fe/H] = -0.24 \pm 0.06$) which is on the same system as the other 
clusters in this 
study.  The age shift described in section 2.1 is also applied to NGC 
2158 giving us an age of $1.6 \pm 0.2$ Gyr. We note in passing that
NGC 2158 was not included as part of our $M_{K}(RC)$ calibration  
because the distance given in Twarog et al. (1997) was determined using 
the magnitude of the RC and not main sequence fitting.  

For the reddening toward NGC 
2158 we can utilize the data in Table 1 to parameterize the
intrinsic color of the red clump ($(J-K)_0$) in terms of the
metal abundance and age.
Figure 8 shows $(J-K)_0$ versus $[Fe/H]$ (left panel) and age (right 
panel) for the clusters in our sample.  
Using the interpolation discussed in section 3.3, we can determine the 
intrinsic color of NGC 2158 given its metallicity 
and age, for which we find $(J-K)_0 = 0.597 \pm 0.003$.  We calculate 
the error in $(J-K)_0$ by determining the uncertainty resulting from 
$\sigma_{age}$ and $\sigma_{[Fe/H]}$ and adding these in quadrature.
Comparing the implied intrinsic color of the red clump
with the apparent color, $(J-K) = 0.821 \pm 0.006$, we find
$E(J-K) = 0.224 \pm 0.007$.  
Converting this to a color excess in the optical regime, we find $E(B-V) 
= 0.43 \pm 0.013$, which is in good agreement with published values 
(e.g. Christian et al. 1985; Twarog et al. 1997). 
The preceding method represents an internally consistent formalism 
which can be utilized to estimate the reddening of a cluster.  

The interpolation on $M_K(RC)$ using only the open cluster
data predicts $M_{K} = -1.64 \pm 0.08$ for NGC 2158. 
Along with $E(B-V) = 0.43$ and the apparent RC K-band magnitude, $K(RC) = 
11.55 \pm 0.02$, we find $(m-M)_{V} = 14.38 \pm 0.09$.  Our distance 
modulus for NGC 2158 agrees within the errors with the main sequence 
fitting modulus of
$(m-M)_{V} = 14.4 \pm 0.2$ found by Christian et al. (1985), but is 
slightly lower than that determined by 
Twarog et al. (1997) of $(m-M)_{V} = 14.5$.

\section{Conclusions}

In this paper, we have sought to establish the K-band absolute 
magnitude of the helium burning red clump stars ($M_K(RC)$) as a 
distance indicator. To facilitate this, we have utilized infrared 
photometry from the 2MASS catalog along with distances, 
metallicities, and ages for 14 open clusters and 2 globular clusters. Our 
sample encompasses an age range from 0.63 Gyr to 12 
Gyr and metallicities from --1.15 to 0.15 dex. Based on an analysis of 
these data, we draw the following conclusions.

1. There is a statistically significant range of $M_K(RC)$ values 
among the star clusters in our sample. In particular, for the 14 open 
clusters, we calculate $\langle$$M_{K}(RC)$$\rangle$$= -1.62$ with a 
standard deviation of 0.21 mag. In contrast, the mean error in these 
$M_K(RC)$ values is 0.13 mag.

2. Upon inspection of figures 5 and 6, we find that for clusters younger 
than $\sim$2 Gyr, $M_K(RC)$ is insensitive to metallicity but shows a 
dependence on age. In contrast, for clusters older than $\sim$2 Gyr,
$M_K(RC)$ is influenced primarily by the metallicity of the population 
and shows little or no dependence on the age.

3. In general, $M_K(RC)$ is less sensitive to age and metallicity than 
$M_I(RC)$ over the parameter range common to both this paper and
Sarajedini (1999) from which the $M_I(RC)$ values are taken.

4. Over comparable metallicity and age ranges, our average $M_K(RC)$ 
value of --1.62 mag is consistent with that 
of Alves (2000) which is based on solar-neighborhood red clump stars
with HIPPARCOS parallaxes.  We also suggest 
that the significant scatter in the Alves (2000) $M_K$ data is 
likely due to a range of ages between $\sim$1.6 and $\sim$4 Gyr
among these stars.

5. The theoretical red clump models based on the formalism of
Girardi et al. (2000) agree reasonably well with our observational data, 
indicating that age plays an important role in determining
$M_K(RC)$ for younger populations while metallicity mainly affects older
populations.

6. Using the K-band absolute magnitude of the red clump, we are able 
to compute the distance to the open cluster
NGC 2158. Adopting an age of $1.6 \pm 0.2$ Gyr and 
$[Fe/H] = -0.24 \pm 0.06$, our calibration yields a 
distance of $(m-M)_{V} = 14.38 \pm 0.09$.

7. When determining distances for star clusters having $-0.5 \leq 
[Fe/H] \leq 0.0$ and $10^{9.2} \leq age \leq 10^{9.9}$, one can ignore 
the interpolation discussed in section 3.3 and simply use $<M_K(RC)>$ = 
$-1.61 \pm 0.04$.

\acknowledgements

We would like to thank Leo Girardi and David Alves for providing us with 
electronic copies of their data.  We would also like to thank Glenn 
Tiede, Ted von Hippel, Barbara Anthony-Twarog, and Bruce Twarog for 
making useful comments that greatly improved the clarity of an early 
manuscript. The comments of an anonymous referee are also greatly
appreciated. A.S. is grateful to the National Science Foundation for
support through grant No. AST-9819768.  This publication makes 
use of data products from the Two Micron All Sky Survey, which is a 
joint project of the University of Massachusetts and the Infrared 
Processing and Analysis Center/California Institute of Technology, 
funded by the National Aeronautics and Space Administration and the 
National Science Foundation.

\clearpage


\begin{figure}
\epsffile{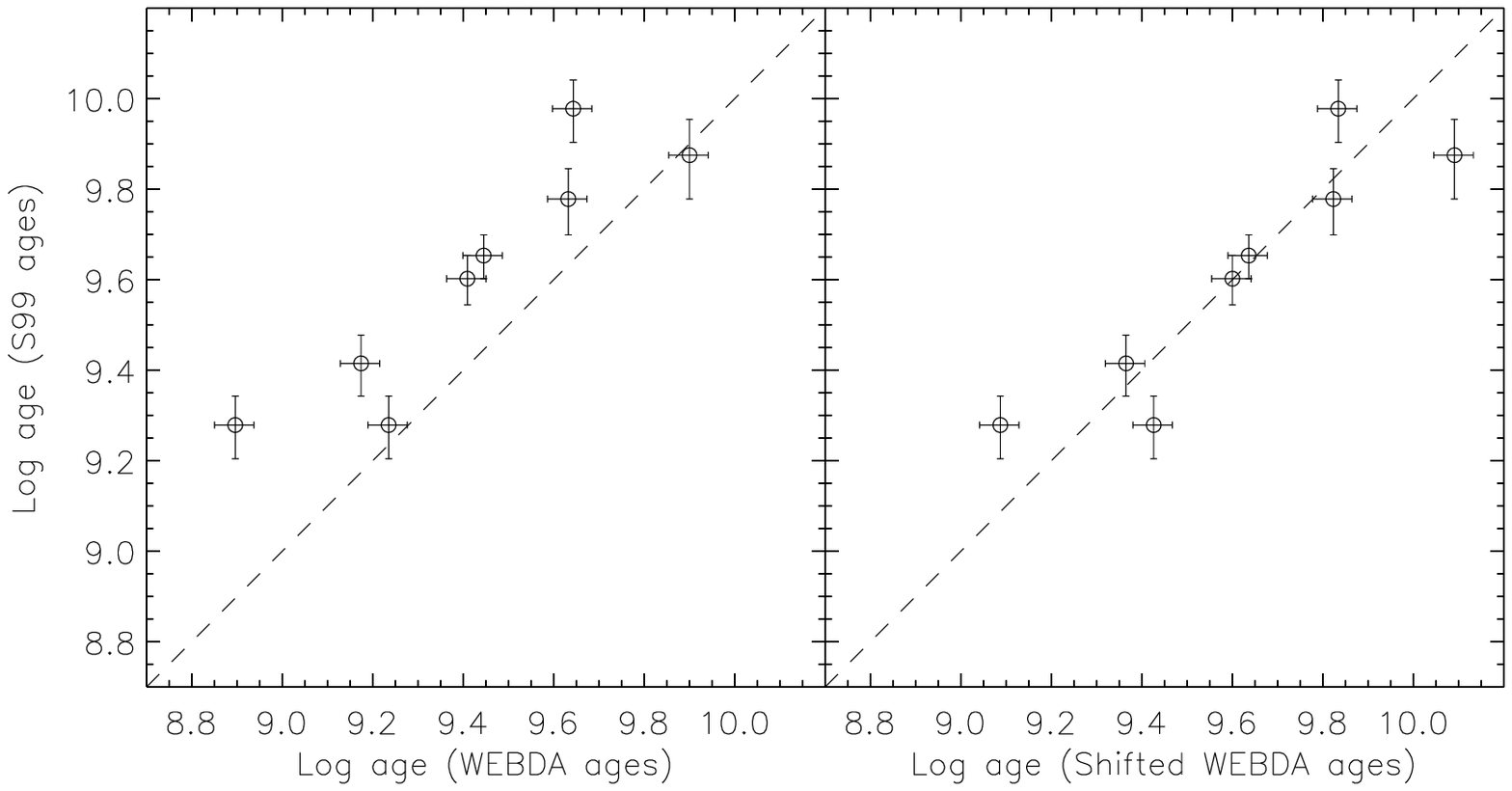}
\figcaption{The left panel shows the ages from WEBDA plotted against 
those from S99.  Due to the systematic difference between the systems, 
we shift the WEBDA ages older (right panel) by $\Delta$ $log (Age) = 
0.191$ 
to place them on the same system as S99.} 
\end{figure}

\begin{figure}
\epsffile{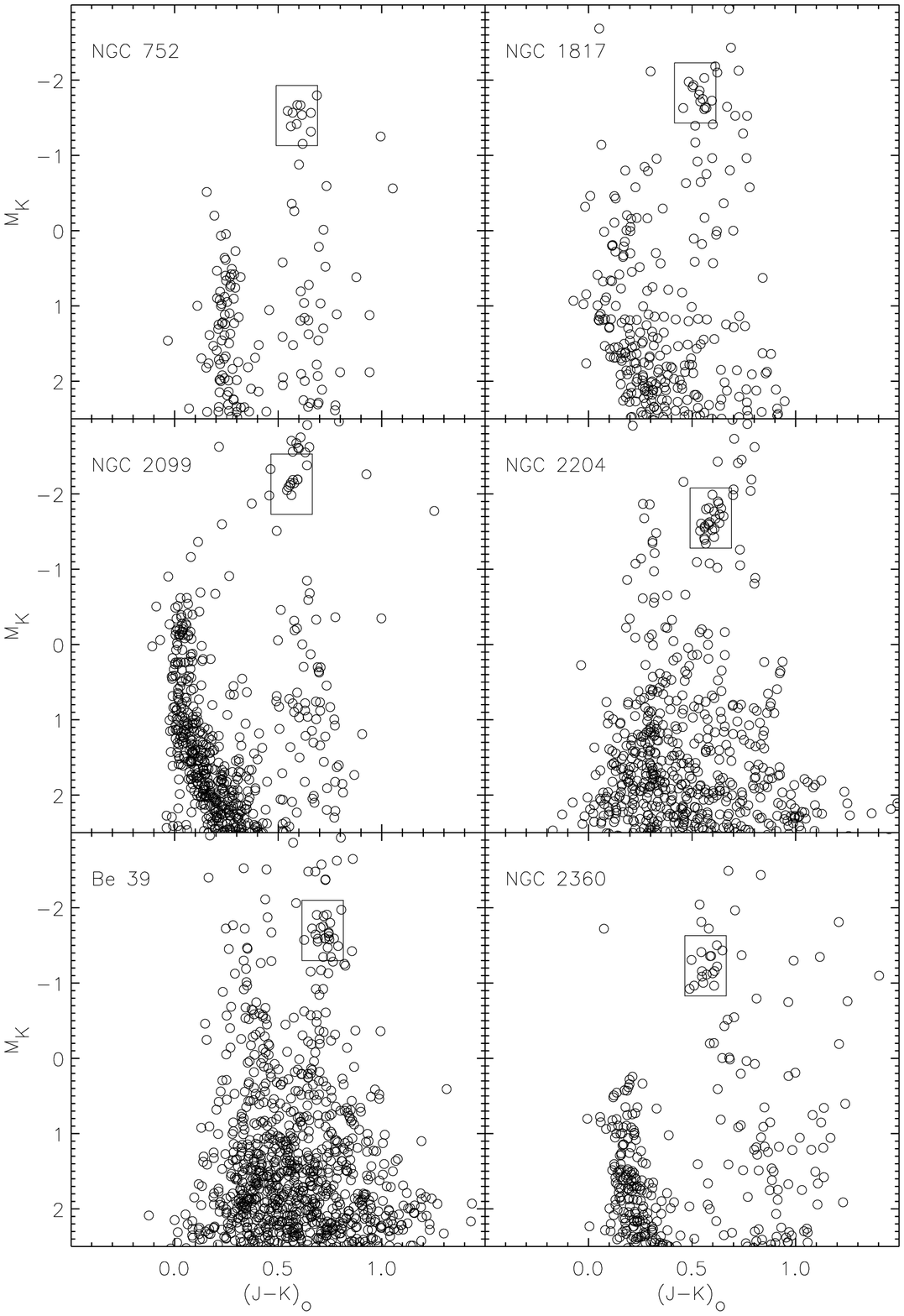}
\end{figure}
\begin{figure}
\epsffile{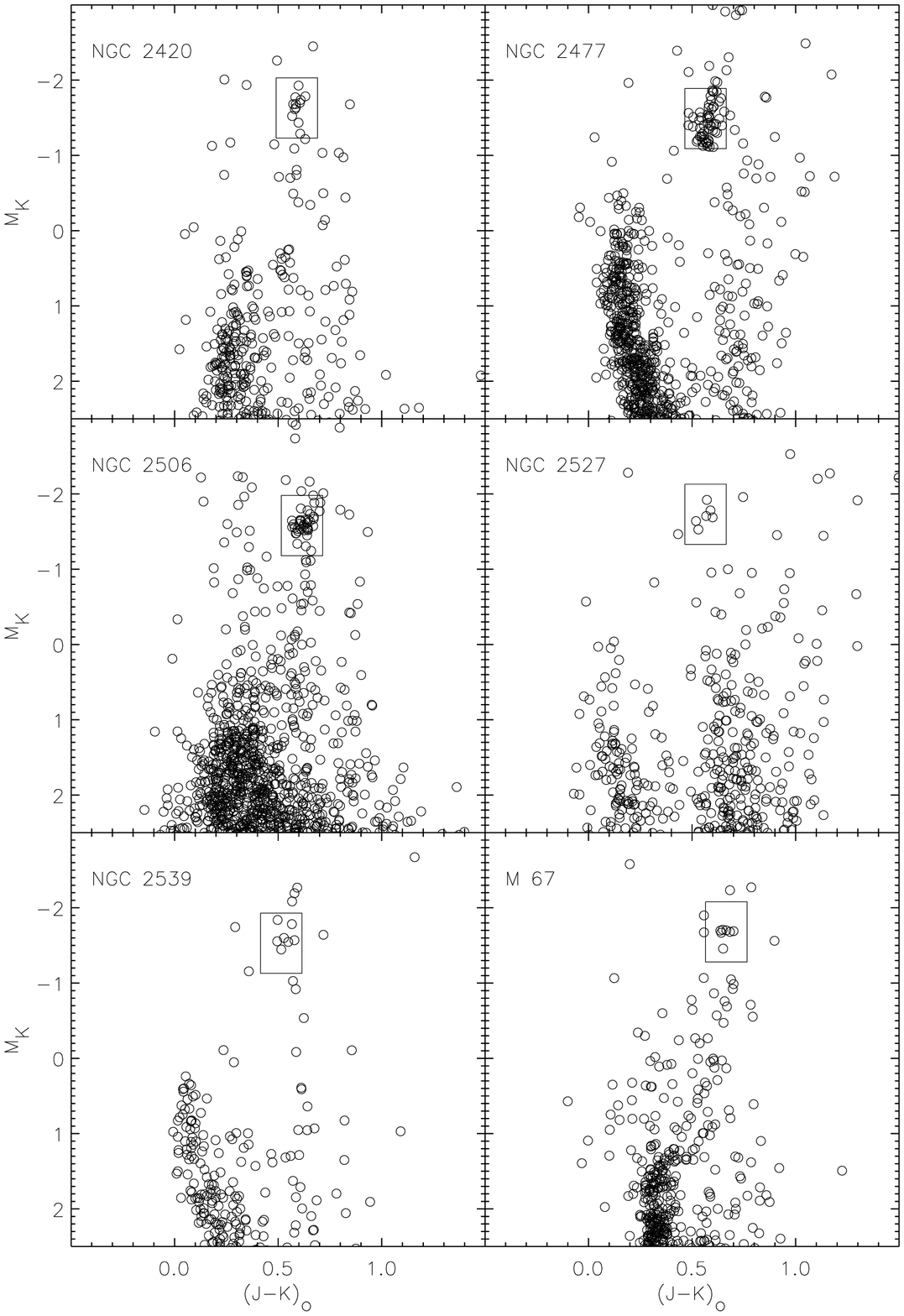}
\end{figure}
\begin{figure}
\epsffile{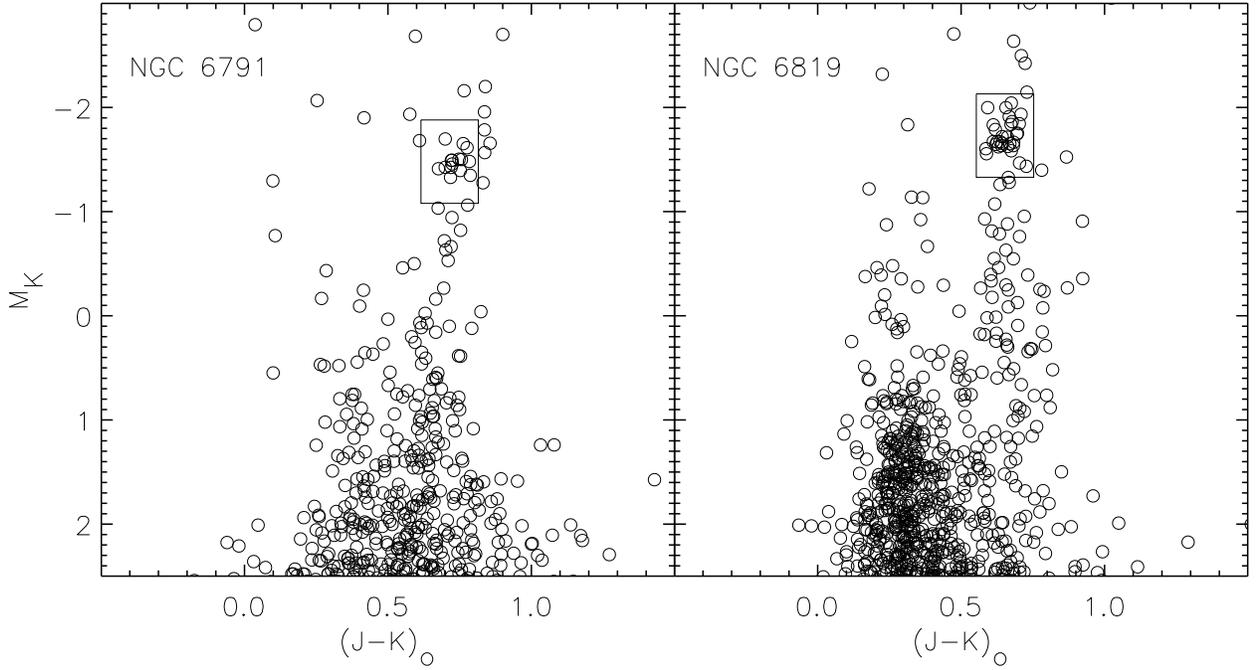}
\figcaption{Infrared color-magnitude diagrams for the 14 open 
clusters in our sample with a box indicating the location of the red 
clump stars.  All stars within the 
box are used in calculating the median K magnitude of the red clump.}
\end{figure}

\begin{figure}
\epsffile{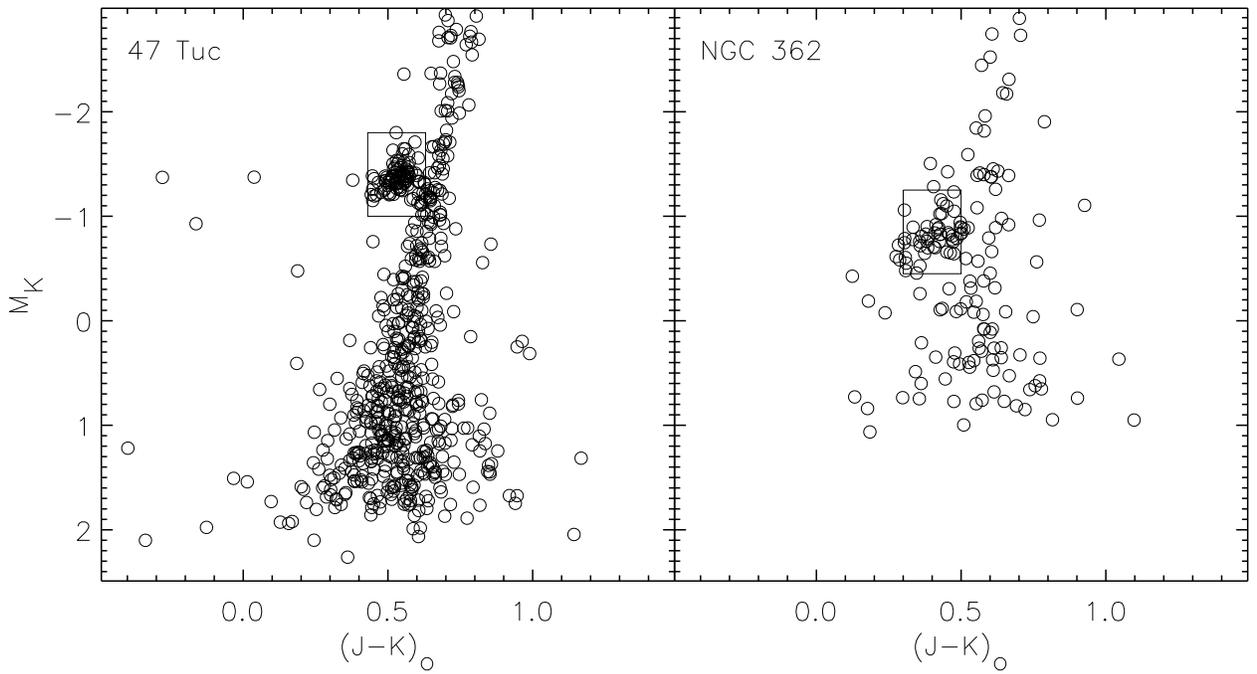}
\figcaption{Same as Fig. 2., but for the globular clusters in our 
sample.}
\end{figure}

\begin{figure}
\epsffile{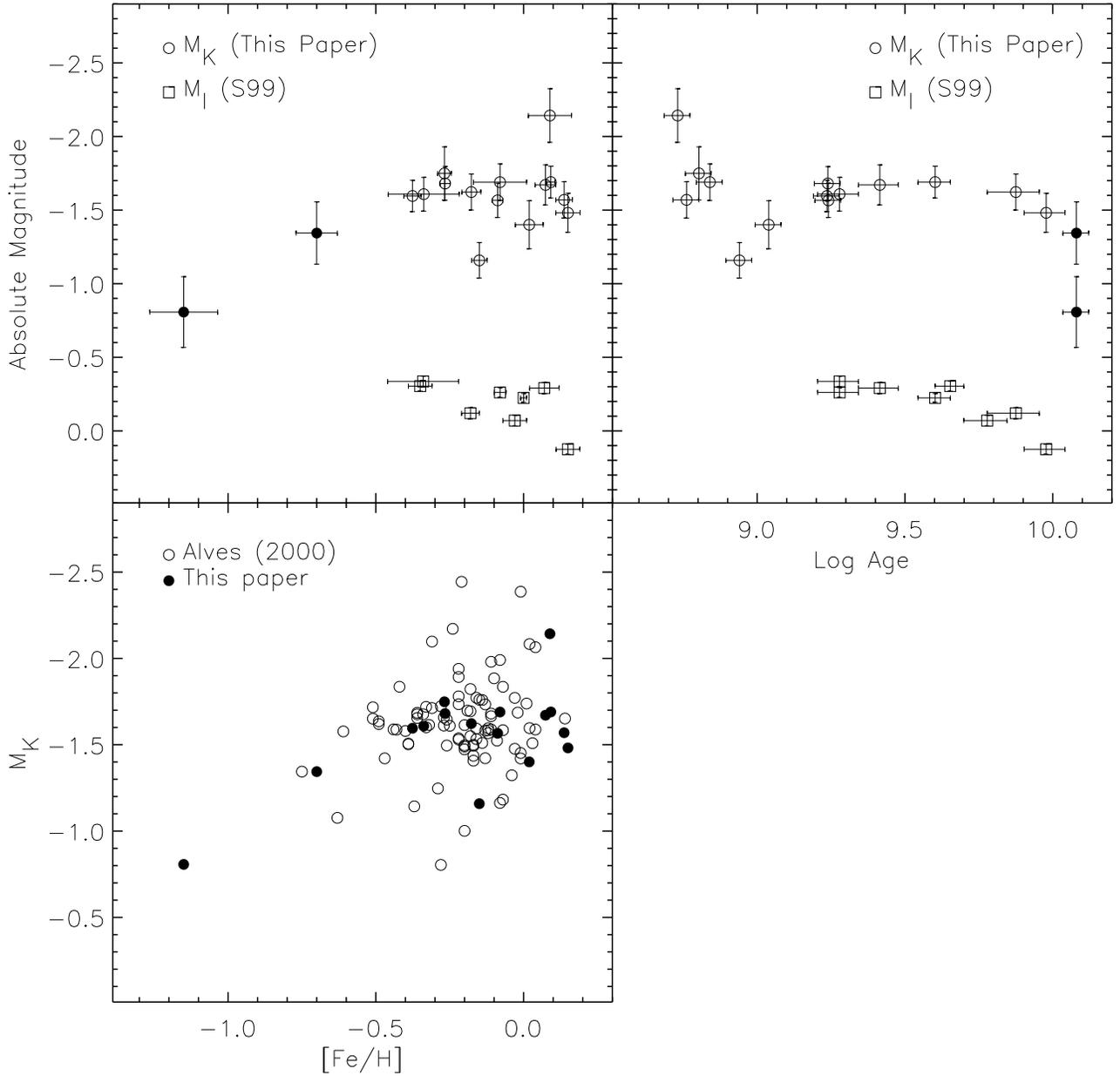}
\figcaption{The upper panels show the variation of red clump absolute
magnitude as a function of age (top left panel) and 
$[Fe/H]$ (top right panel). The open circles represent K-band absolute
magnitudes ($M_K(RC)$) for the 14 open clusters while the filled 
circles signify $M_K(RC)$ values for the two globular clusters 
in the present sample. The open squares designate $M_I(RC)$ values 
from Sarajedini (1999).  In the bottom panel, $M_K$ 
for HIPPARCOS Solar neighborhood red clump stars from Alves (2000) 
(open circles) are compared with $M_K(RC)$ for clusters in the present paper 
(filled circles). These two data sets show remarkable agreement in 
their mean K-band magnitudes.}
\end{figure}

\begin{figure}
\epsffile{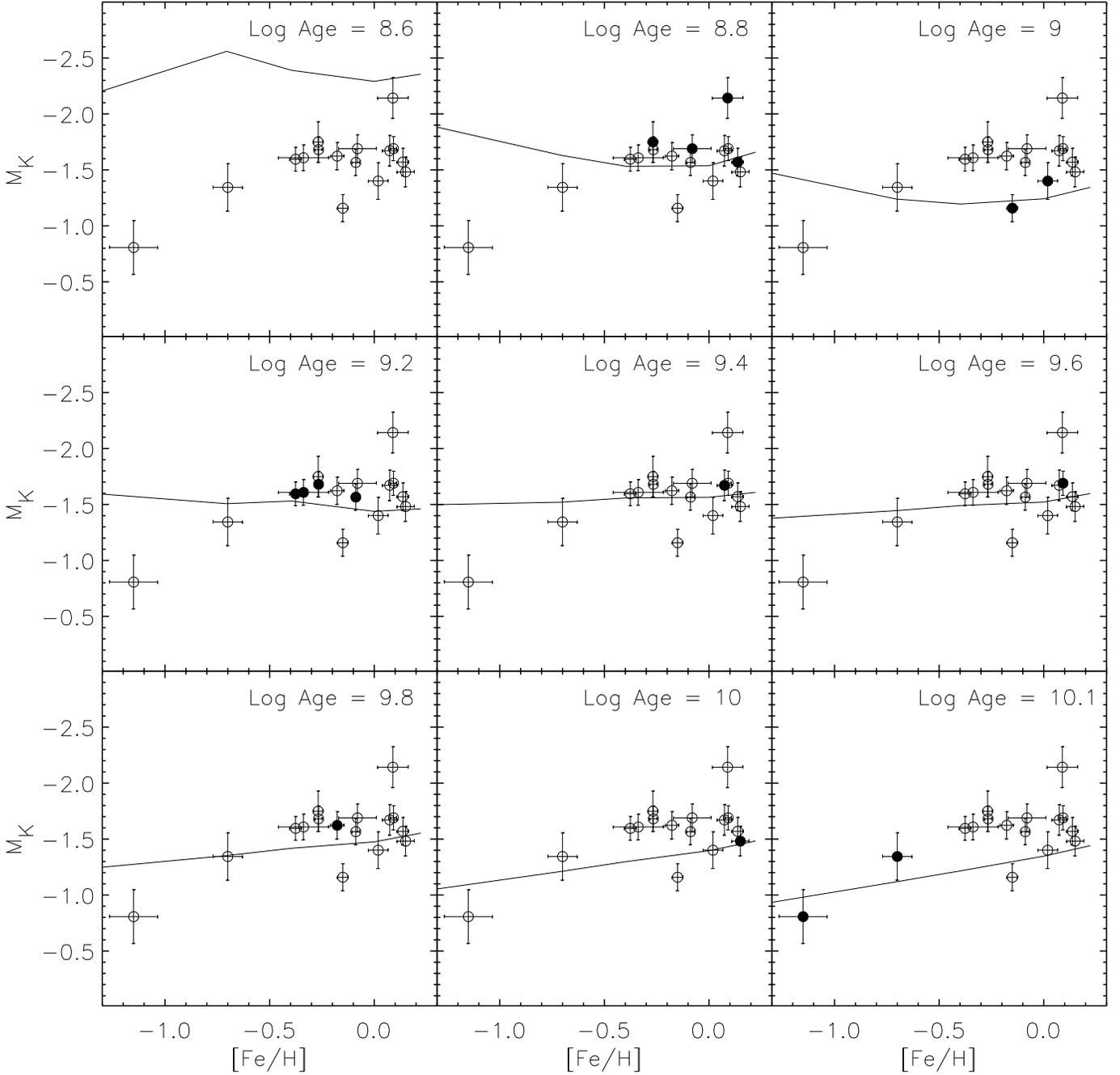}
\figcaption{The observed variation of $M_K(RC)$ with the logarithm of 
the age as compared with the predictions of theoretical models (Girardi et 
al. 2000) for the indicated metallicities. The filled circles represent 
clusters with ages that are within $\pm$0.1 dex of the model age in each
panel. For the upper left and lower right panels, 
the filled circles represent clusters with $log (Age) \leq 8.7$ and 
$log (Age) \geq 10.05$, respectively. The remaining clusters 
in each panel are marked by open circles.}
\end{figure}

\begin{figure}
\epsffile{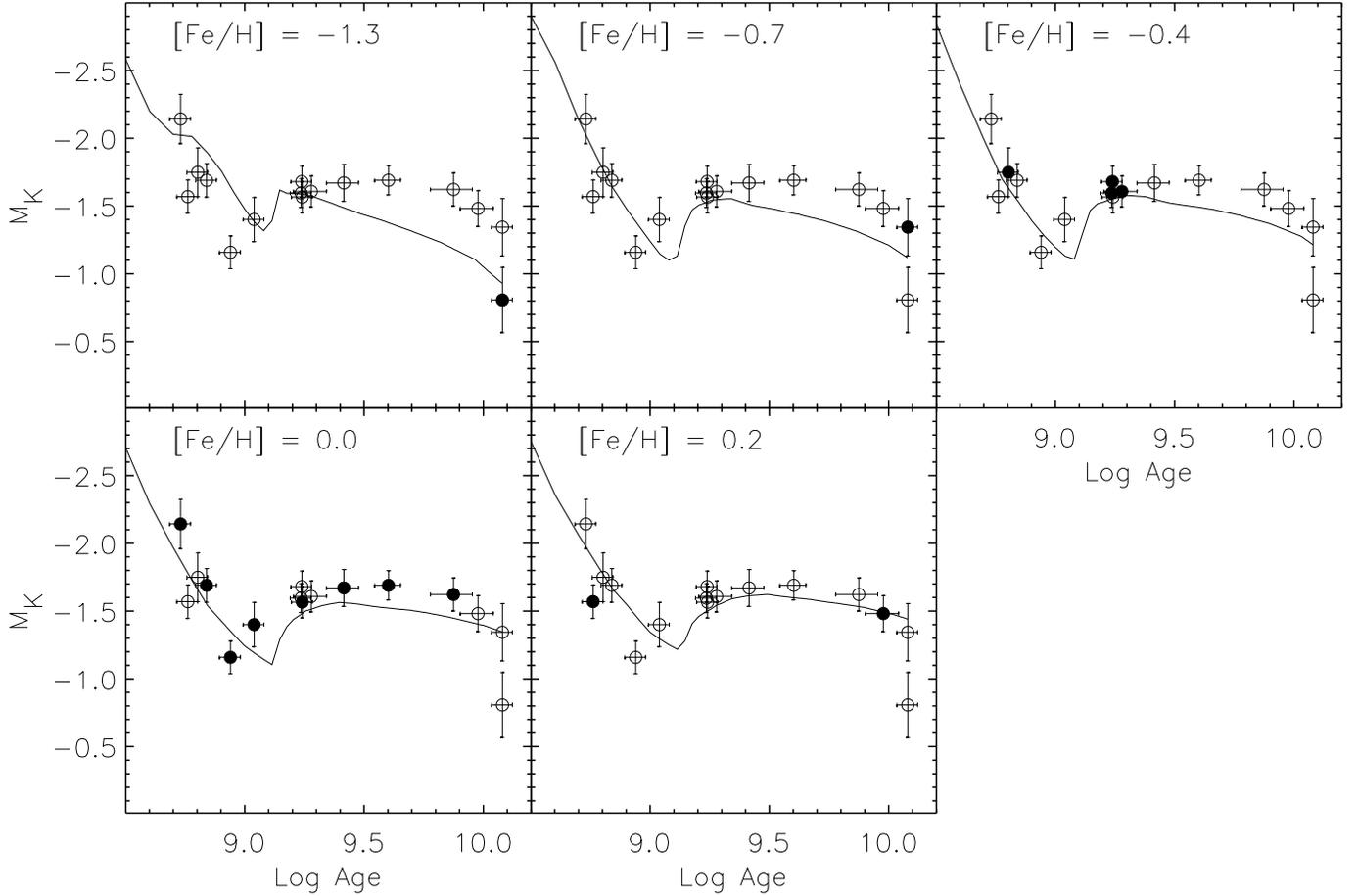}
\figcaption{The observed variation of $M_K(RC)$ with metal abundance 
as compared with the predictions of theoretical models (Girardi et 
al. 2000) for specific ages. The filled circles denote the clusters 
with $[Fe/H]_{min} \leq [Fe/H] \leq [Fe/H]_{max}$, where $[Fe/H]_{min}$ 
and $[Fe/H]_{max}$ are halfway between the model shown and the next 
lower and next higher metallicity models, respectively.  
For the models of $[Fe/H] = -1.3$ and $[Fe/H] = 0.2$, clusters having 
$[Fe/H] \leq -1.0$ and $[Fe/H] \geq 0.1$, respectively, are marked with 
filled circles.  The remaining clusters in each panel are shown by open 
circles.}
\end{figure}

\begin{figure}
\epsffile{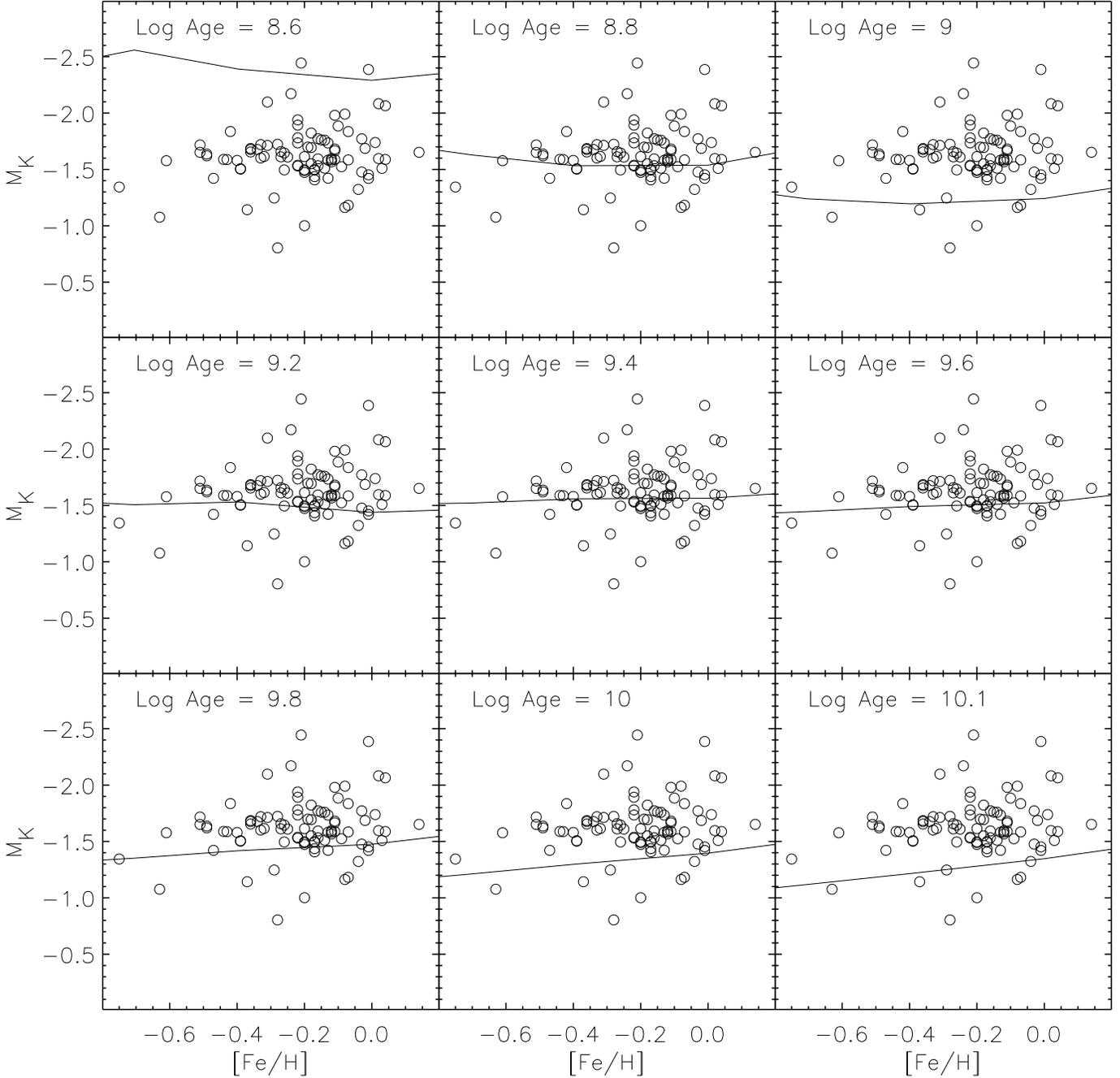}
\figcaption{The K-band absolute magnitude of solar-neighborhood red
clump stars with HIPPARCOS parallaxes versus their metallicities.
The solid lines represent the predictions of theoretical models 
constructed by Girardi et al. (2000).  The models suggest that the vertical 
spread in $M_K$ 
can be explained by variations in the ages of the stars.}
\end{figure}

\begin{figure}
\epsffile{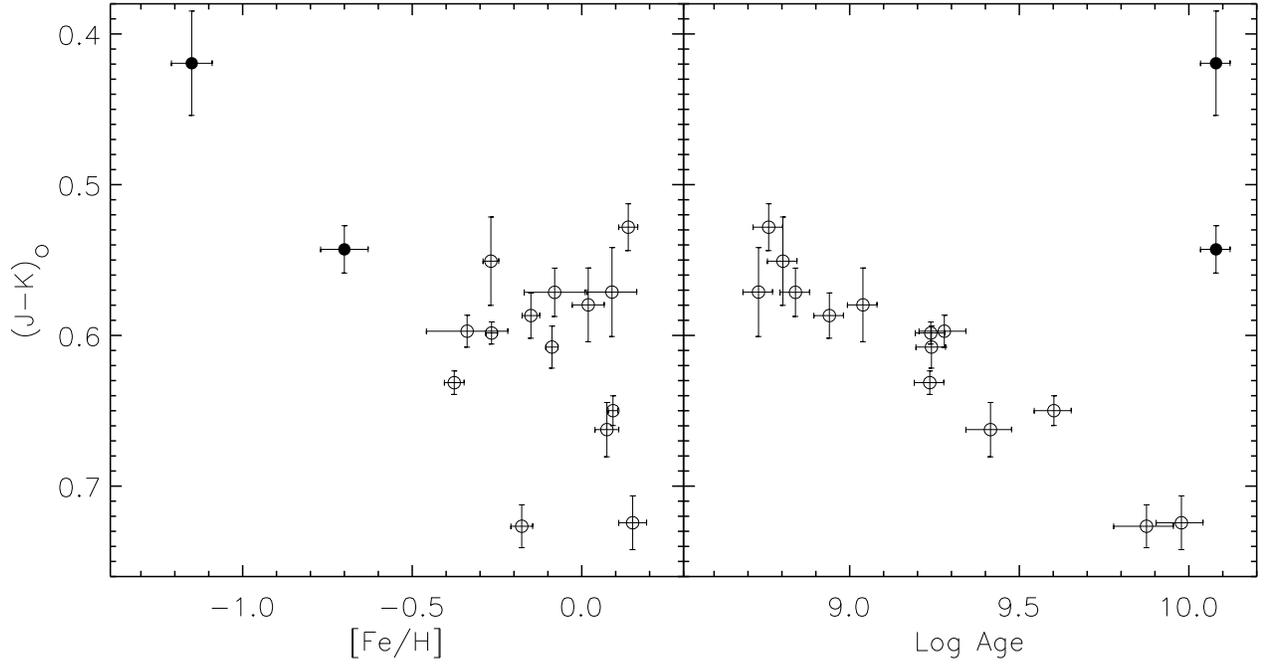}
\figcaption{The left panel shows the intrinsic color of the RC plotted 
as a function of $[Fe/H]$ and the right panel shows it plotted as a 
function of age.  The open circles represent the open clusters while the 
filled circles are the globulars.}
\end{figure}



\begin{deluxetable}{lccccccccc}
\tabletypesize{\scriptsize}
\tablecaption{Open Cluster Information. \label{tbl-1}}
\tablewidth{0pt}
\tablehead{
\colhead{Name} & 
\colhead{Log Age}   & 
\colhead{$(m-M)_V$\tablenotemark{a}}   &
\colhead{$E(B-V)$\tablenotemark{a}} & 
\colhead{$[Fe/H]$\tablenotemark{a}}  & 
\colhead{$\sigma ([Fe/H])$\tablenotemark{a}} &
\colhead{$M_K$} & 
\colhead{$\sigma (M_K)$} &
\colhead{$(J-K)_{o}$} &
\colhead{$\sigma (J-K)_{o}$}
}
\startdata
NGC 752 &9.24 &8.35 &0.04 &$-$0.088 &0.018 &$-$1.566 &0.116 &0.608 
   &0.014\\
NGC 1817 &8.80 &12.15 &0.26 &$-$0.268 &0.023 &$-$1.749 &0.180 &0.551 
   &0.029\\
NGC 2099 &8.73 &11.55 &0.27 &0.089 &0.073 &$-$2.143 &0.182 &0.571 
   &0.030\\
NGC 2204 &9.28\tablenotemark{b} &13.30 &0.08 &$-$0.338 &0.120 &$-$1.608
   &0.115 &0.597 &0.011\\
Be 39 &9.88\tablenotemark{b} &13.50 &0.11 &$-$0.177 &0.032 &$-$1.623
   &0.121 &0.727 &0.014\\
NGC 2360 &8.94 &10.35 &0.09 &$-$0.150 &0.026 &$-$1.159 &0.121 &0.587 
   &0.015\\
NGC 2420 &9.24 &12.10 &0.05 &$-$0.266 &0.017 &$-$1.681 &0.115 &0.598 
   &0.007\\
NGC 2477 &9.04 &11.55 &0.23 &0.019 &0.047 &$-$1.401 &0.164 &0.580 
   &0.024\\
NGC 2506 &9.24 &12.60 &0.05 &$-$0.376 &0.029 &$-$1.596 &0.106 &0.631 
   &0.008\\
NGC 2527 &8.84 &9.30 &0.09 &$-$0.080 &0.090 &$-$1.690 &0.124 &0.571 
   &0.016\\
NGC 2539 &8.76 &10.75 &0.09 &0.137 &0.028 &$-$1.570 &0.123 &0.528 
   &0.016\\
M 67 &9.60\tablenotemark{b} &9.80 &0.04 &0.092 &0.014 &$-$1.690 &0.108 
   &0.650 &0.010\\
NGC 6791 &9.98\tablenotemark{b} &13.40 &0.15 &0.150 &0.041 &$-$1.482 
   &0.132 &0.724 &0.018\\
NGC 6819 &9.42\tablenotemark{b} &12.44\tablenotemark{b}
   &0.16\tablenotemark{b} &0.074 &0.035 &$-$1.671 &0.136 &0.663 &0.018\\
47 Tuc &10.08 &13.45\tablenotemark{c} &0.044\tablenotemark{c} 
   &$-$0.70\tablenotemark{d} &0.07\tablenotemark{d} &$-1.344$ &0.211 
   &0.543 &0.016\\
NGC 362 &10.08 &14.70\tablenotemark{c} &0.048\tablenotemark{c} 
   &$-$1.15\tablenotemark{d} &0.06\tablenotemark{d} &$-0.807$ &0.241 
   &0.419 &0.035\\
\enddata


\tablenotetext{a}{From Twarog et al. (1997) unless otherwise noted}
\tablenotetext{b}{From Sarajedini (1999)}
\tablenotetext{c}{See section 2.2}
\tablenotetext{d}{From Carretta and Gratton (1997)}

\end{deluxetable}

\end{document}